\documentclass[fleqn,10pt]{wlscirep}
\usepackage[utf8]{inputenc}
\usepackage[T1]{fontenc}
\usepackage{textgreek}
\title{Preparing pure $^{43}$Ca$^+$ samples in an ion trap with photoionization and parametric excitations}

\author[1,2]{C.-H. Kuo}
\author[1,2]{Y.-C. Hsiao}
\author[1,2]{C.-Y. Jhang}
\author[1,2]{Y.-D. Chen}
\author[1,2,*]{S. Tung}
\affil[1]{ Department of Physics, National Tsing Hua University, Hsinchu 30013, Taiwan}
\affil[2]{Center for Quantum  Science and Technology, Hsinchu 30013, Taiwan}

\affil[*]{stung@phys.nthu.edu.tw}

\begin{abstract}
We present a practical scheme for the efficient preparation of laser-cooled $^{43}$Ca$^+$ ions in an ion trap. Our approach integrates two well-established methods: isotope-selective photoionization and isotope-specific parametric excitation. Drawing inspiration from the individual merits of each method, we have successfully integrated these techniques to prepare extended chains of $^{43}$Ca$^+$ ions, overcoming the challenge posed by their low natural abundance of 0.135\% in a natural source. Furthermore, we explore the subtleties of our scheme, focusing on the influence of different factors on the purification process. Our investigation contributes to a broader understanding of the technique and highlights the adaptability of established methods in addressing specific isotopic challenges.
\end{abstract}
\begin{document}

\flushbottom
\maketitle
%
%
\thispagestyle{empty}


\section*{Introduction}

Ion traps have a plethora of applications, including quantum information processing \cite{PhysRep.469.155, RevModPhys.82.1209, ApplPhysRev.6.021314}, optical clocks \cite{RMP87.637, Phys.Rev.Lett.116.063001, Appl.Phys.B.123.166, Phys.Rev.Lett.123.033201}, and precision measurements \cite{PRL.115.053003, PRA.128.033202, PRL.125.123002, PRL.128.033202}. In these applications, the preparation of isotope-specific atomic ions in an ion trap is usually the first step, and this task becomes challenging when the desired isotope has a low natural abundance. To address this issue, different techniques, such as photoionization \cite{Appl.Phys.B71.207, Appl.Phys.B73.861, PhysRevA.69.012711, PhysRevA.75.053404, ApplPhysB.87.411,  ApplPhysB.108.159} and laser ablation \cite{OptExp.27.33907, PhysRevA.105.033102}, have been utilized to prepare isotope-specific ions. Yet, in these processes, ions of unwanted isotopes may still be generated. To remove these isotopes, techniques such as laser manipulation  \cite{PhysRevA.69.012711, J.Phys.B.45.165008}, nonlinear resonances \cite{ApplPhysB.62.511, ApplPhysB.65.57, ApplPhysB.89.195, InternationalJ.MassSpectrum.279.163, ApplPhysB.103.339}, and parametric excitation \cite{ApplPhysB.70.867, PhysRevA.66.063414, ApplPhysB.126.176} have been employed. 

Calcium (Ca) is a prevalent element in trapped-ion research, best known for its role in quantum information processing \cite{Nature.438.643, PhysRevLett.106.130506, Science.334.57, PhysRevLett.117.060504, Nature.555.75, Nature.528.384, PhysRevLett.113.220501, PhysRevLett.117.140501, NaturePhysics.18.296, Nature.605.675}. With optical transitions at convenient wavelengths, calcium ions mitigate the complexity of optical manipulations required for many applications. Notably, a compact quantum computing demonstrator featuring calcium ions has been recently developed \cite{PRXquantum2.020343}. Furthermore, the presence of metastable states with exceptionally long lifetimes makes the element suitable for optical clocks \cite{PhysRevLett.102.023002, PhysRevLett.116.013001, PhysRevApplied.17.034041}. Finally, single calcium ions can also serve as single photon sources \cite{NewJPhys.11.103004, NewJPhys.15.055005, NewJPhys.18.093038}, a crucial component in quantum communication and cryptography \cite{RepProgPhys.68.1129, RevSciInstrum.82.071101}. These applications underscore the versatility of trapped calcium ions in quantum technological applications.

$^{43}$Ca stands out as the only stable calcium isotope with non-zero nuclear spin. The resultant hyperfine levels can be used for hyperfine qubits characterized by extended coherence times. However, taking advantage of the unique properties of $^{43}\textrm{Ca}^+$ is no trivial task, given its scarce natural abundance of 0.135\%, especially when compared to the predominant $^{40}\textrm{Ca}^+$, which constitutes nearly 97\% of natural calcium. While a limited number of research groups have turned to enriched sources of $^{43}$Ca to enhance trap loading selectivity, the availability of these sources is limited. Isotope purification represents an alternative strategy that selectively removes ions of undesired isotopes from mixtures. 

Laser manipulation offers a method for ion purification, employing blue-detuned or resonant lasers to heat and eject unwanted isotopes from the trap, while red-detuned lasers cool the desired isotope \cite{PhysRevA.69.012711, J.Phys.B.45.165008}. In their notable work, ref.~\cite{PhysRevA.69.012711}, Lucas, D. M. {\it et al.} demonstrated the potential of this method for purifying $^{43}\textrm{Ca}^+$ ions. However, the effectiveness of laser-assisted purification is highly dependent on the spectrum of the atomic species and the laser parameters.

Another method of purification employs parametric excitations. It has been demonstrated that resonant parametric excitations are effective for removing $^{25}$Mg$^+$ and $^{26}$Mg$^+$, which possess respective natural abundances of 10\% and 11\% \cite{ApplPhysB.70.867}. Recently, same method is adopted to purify $^{138}$Ba$^+$ \cite{ApplPhysB.126.176}. However, the combined efficacy of parametric excitations and resonance ionization in isolating rare isotopes, such as $^{43}\textrm{Ca}^{+}$, remains a topic of investigation. In this article, we report on a procedure that combines isotope-selective photoionization and isotope-specific parametric excitations to achieve efficient preparation of $^{43}\text{Ca}^{+}$ ions. We also demonstrate that this technique can be readily extended to any selected calcium isotopes. 


\section*{Experimental setup}\label{experiment}

In our experiment, we utilize a blade-type ion trap inspired by the design presented in ref.~\cite{Gulde}. The trap consists of four blade electrodes that provide confinement in the radial direction (x-y plane) and two endcap electrodes that enable confinement along the axial direction (z axis). An rf voltage source capable of reaching an amplitude of 200V and operating at a frequency of 44.2 MHz is applied to a diagonal pair of the four blade electrodes, while the opposite pair is grounded. Additionally, a dc voltage of approximately 10 V is applied to the endcaps. Within the trap, ions are laser-cooled and positioned 0.5 mm away from the blade electrodes and 1.2 mm away from the endcap electrodes. With these experimental parameters, we have recorded radial and axial secular frequencies $(\omega_r, \omega_z)=2\pi\times(685, 346)$ kHz for $^{40}$Ca$^{+}$.

\begin{figure}[h]
	\centering
	\includegraphics[width=0.5\textwidth]{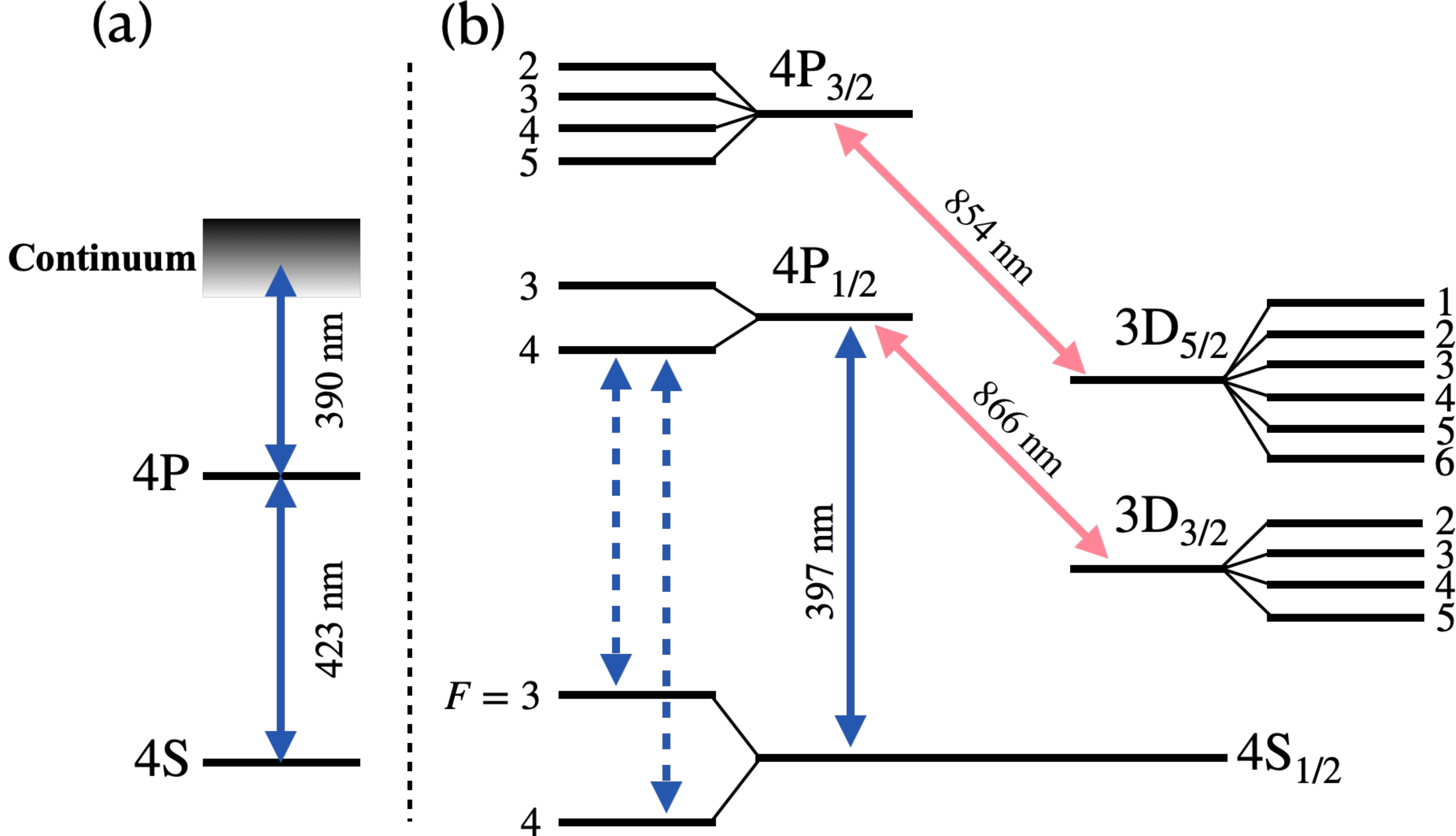}
	\caption{Energy level diagrams for (a) neutral calcium and (b) $\textrm{Ca}^{+}$. The hyperfine structure belongs to $^{43}\textrm{Ca}^{+}$, and the arrows indicate the transitions used in the experiment.}
	\vspace{-4mm}
	\label{Fig1}
\end{figure}

In our experiments, we utilize three external cavity diode lasers for the laser cooling of calcium ions. These lasers operate at wavelengths of 397 nm, 866 nm, and 854 nm, as illustrated in Fig.~\ref{Fig1}. Their frequencies are stabilized using a wavemeter. For $^{40}$Ca$^+$ ions, the frequency of the 397-nm laser is set to 30 MHz below the $4\text{S}_{1/2} \rightarrow 4\text{P}_{1/2}$ transition. In contrast, the 866-nm and 854-nm lasers are on resonance with the $3\text{D}_{3/2} \rightarrow 4\text{P}_{1/2}$ and $3\text{D}_{5/2} \rightarrow 4\text{P}_{3/2}$ transitions, respectively. For the $^{42}$Ca$^+$ and $^{44}$Ca$^+$ ions, we adjust the laser frequencies to compensate isotope shifts and maintain detuning settings similar to those optimized for $^{40}$Ca$^+$.

$^{43}$Ca$^+$ requires a unique laser configuration because of its hyperfine structure. Here, we lock the 397-nm laser to the $4\text{S}_{1/2}$(F=4) $\rightarrow$ $4\text{P}_{1/2}$(F$^{\prime}$=4) transition, detuned by -35 MHz. This laser's output is then sent through a phase modulator to generate a 3.226 GHz red sideband, responsible for driving the $4\text{S}_{1/2}$(F=3) $\rightarrow$ $4\text{P}_{1/2}$(F$^{\prime}$=4) transition. We also stabilize the 866-nm laser to operate at 345.996436 THz. This laser's output is split into three portions, each subjected to unique frequency shifts of 110, 350 and 500 MHz. These adjusted outputs are tailored to induce transitions between the hyperfine levels of the $3\text{D}_{3/2}$ and $4\text{P}_{1/2}$ states, facilitating the repumping of ions to the ground state.

The process to prepare ultracold ions starts with the photoionization of neutral calcium atoms. These neutral atoms, emitted from a heated effusive oven, are channeled into the trap where they undergo ionization through a two-photon mechanism \cite{PhysRevA.69.012711}. Initially, a photon from a 423-nm laser excites a neutral Ca atom from the $4\text{S}$ to the $4\text{P}$ state. Subsequently, another photon from an LED light source with a wavelength of approximately 390 nm ionizes the atom. It is noteworthy that the intrinsic linewidth of the $4\text{S}\rightarrow4\text{P}$ transition is $2\pi\times35$ MHz. This linewidth is considerably narrower than the isotope shifts observed in calcium, thereby enabling isotope-specific ionization. However, Doppler broadening of the transitions severely impedes  the selectivity. To address this broadening, we have positioned a pinhole between the oven outlet and the trap center, which refines the transverse velocity distribution of the atomic beam. Furthermore, the 423-nm laser is aligned perpendicularly to the atomic beam, offering additional mitigation against this broadening.

We employ fluorescence imaging to detect ions in our experiments. Specifically, the fluorescence emitted by the ions is collected by an objective lens and subsequently imaged onto a camera through a lens tube. The objective lens, featuring a numerical aperture of 0.13 and a working distance of 48.5 mm, collaborates with the lens tube to achieve a 10x magnification. To ensure only the fluorescence from the 4S$_{1/2} \rightarrow$ 4P$_{1/2}$ transition (397 nm) is detected, an interference filter is integrated to exclude near-infrared radiation. The advantage of fluorescence imaging is twofold: it allows for precise calibration of fluorescence intensity based on the ion images and enables the detection of dark ions within ion crystals, which manifest as dark gaps in an otherwise regular ionic array.
\vspace{-4mm}

\begin{figure}[t]
	\centering
	\includegraphics[width=0.45\textwidth]{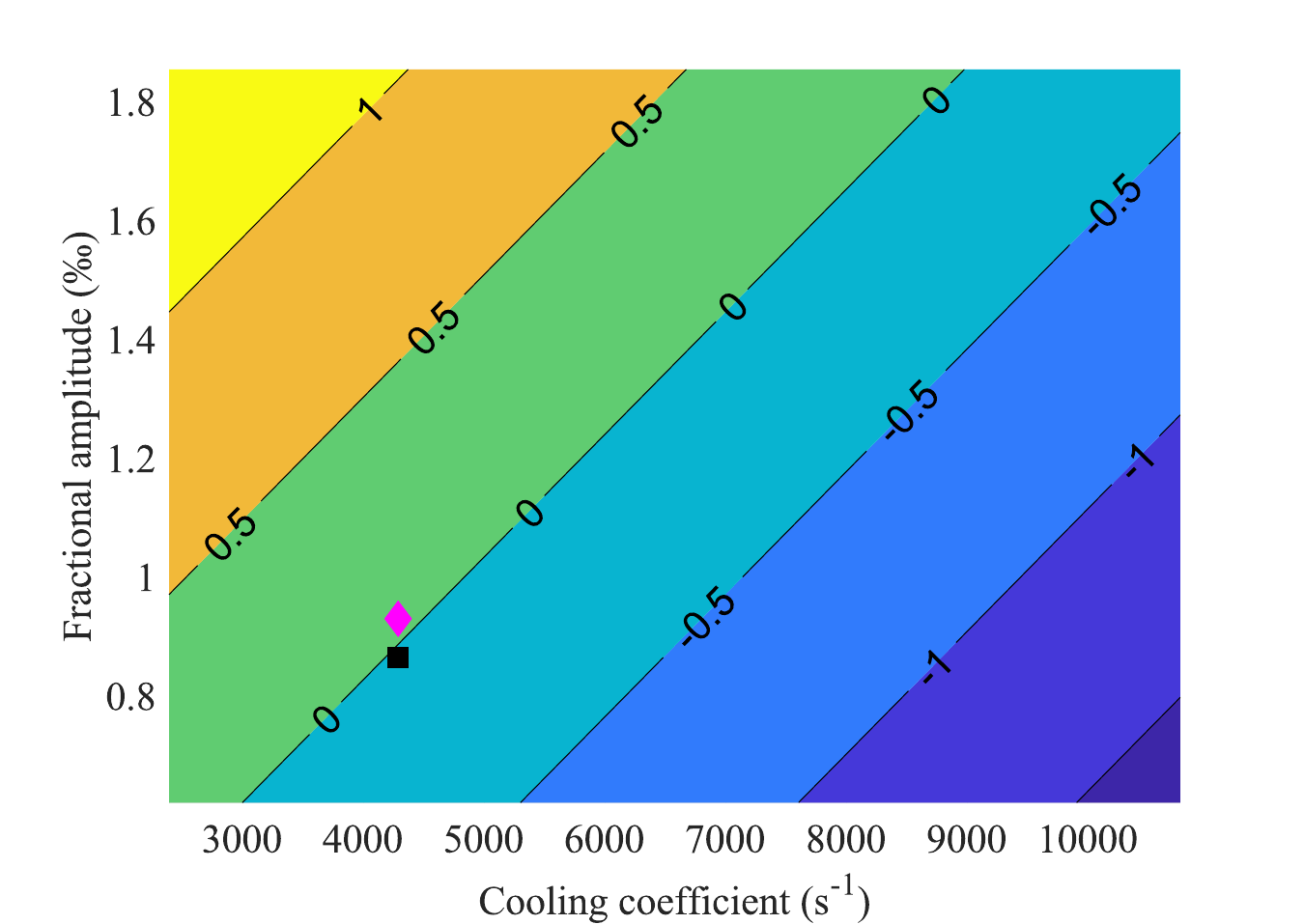}
	\caption{Logarithmic ratio of final to initial motional amplitudes, $\log[ y(\Delta t)/y(0)]$, of a single $^{40}\textrm{Ca}^+$ ion for a range of cooling coefficients $\beta$ and fractional amplitude $\epsilon_0$, calculated from Eq.~(\ref{EOM2}). In the calculation, we fix $a = -1.216\times10^{-4}$ and $q = 4.649\times10^{-2}$, and choose y(0) = 0.25 $\mu$m. The modulation frequency is set to 1370 kHz, corresponding to the resonant frequency of the radial parametric excitations. The graph highlights the balance between parametric heating and laser cooling, with dominant laser cooling leading to a diminished motional amplitude and dominant parametric drive leading to increased amplitude. The diamond and square symbols respectively correspond to the parameters of two different data points in Fig.~\ref{Fig4}(a), illustrating a heating dominated and a cooling dominated scenario.}
	\label{Fig2}
\end{figure}

\section*{Parametric excitation}\label{paramheat}

The dynamics of an ion trapped within an ion trap can be effectively described by the Mathieu equations, as demonstrated in ref.~\cite{RevModPhys.75.281}:

\begin{equation}
\frac{d^2u}{d\xi^2}+\Big[a_u+2q_u\text{cos}(2\xi)\Big]u=0.
\label{EOM}
\end{equation}
In this representation, $\xi$ is defined as $\Omega t/2$, with $\Omega$ denoting the angular frequency of the rf field, and $u$ represents any of the spatial coordinates: $x$, $y$, or $z$. The parameters $a$ and $q$ dictate the ion's stability within the trap. In our experimental setup, we have $a = a_x = a_y = -a_z/2$ and $q = q_x=-q_y$ with        
\begin{equation} \label{parameters}
a=-\frac{4eV_{\text{dc}}\alpha}{mz_{0}^{2}\Omega^2},\quad
q=\frac{2eV_{\text{rf}}\alpha'}{mr_{0}^{2}\Omega^2}.
\end{equation}
Here, $e$ and $m$ represent the ion's charge and mass, respectively. $V_{\text{dc}}$ is the potential applied to the trap's dc electrodes, while $V_{\text{rf}}$ corresponds to the potential of rf electrodes. The terms $z_{0}$ and $r_0$ symbolize the respective distances between the ion and the dc and rf electrodes. $\alpha$ and $\alpha'$ are geometric factors of the trap assembly.

\begin{figure}
    \centering
	\includegraphics[width=0.5\textwidth]{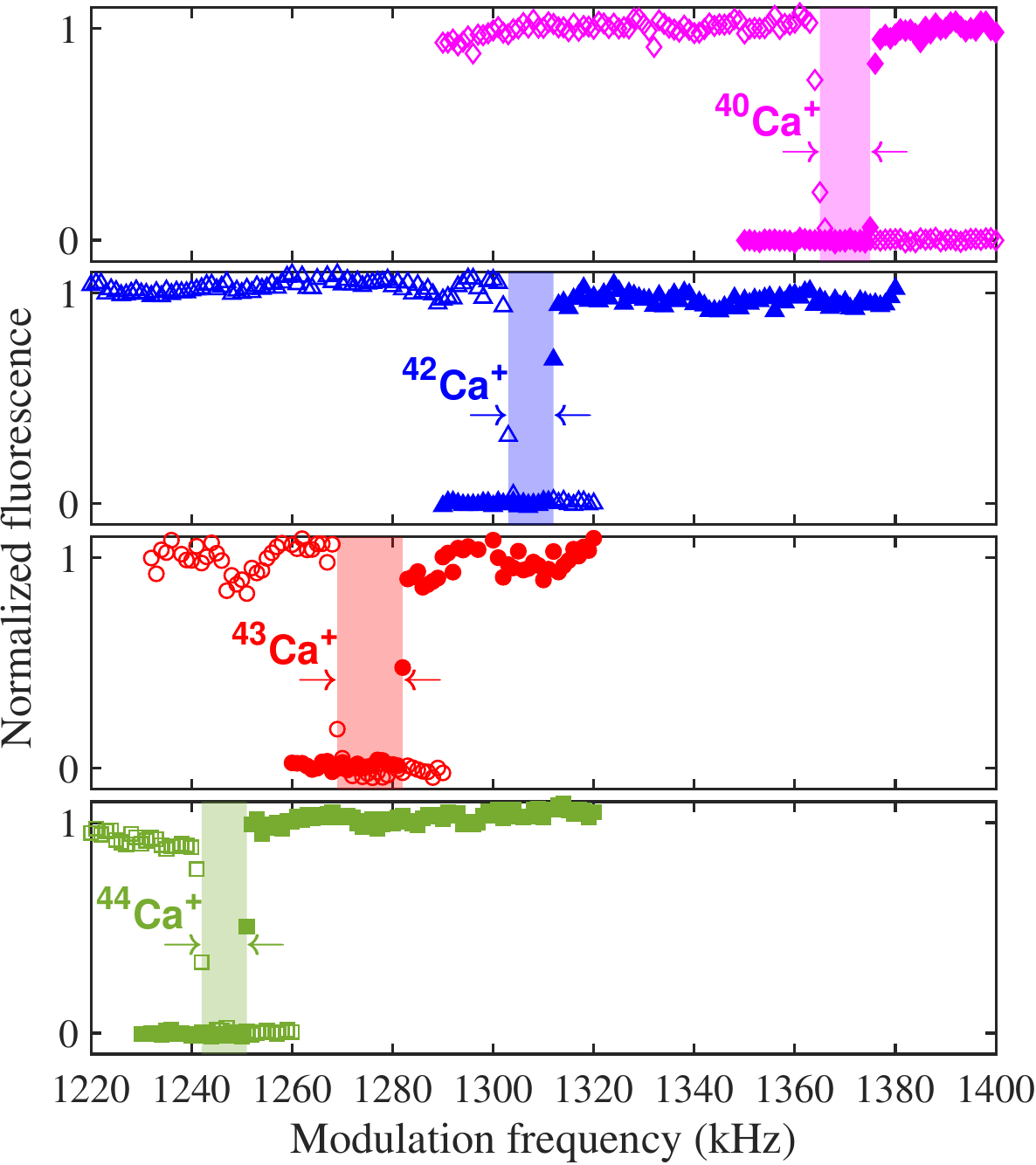}
	\caption{Normalized fluorescence as a function of the RF modulation frequency for the parametric ejection of different calcium isotopes. Each panel shows the response for $^{40}$Ca$^+$(top, diamonds), $^{42}$Ca$^+$(second, triangles), $^{43}$Ca$^+$(third, circles), and $^{44}$Ca$^+$(bottom, squares). Open symbols denote data collected during an ascending frequency sweep, and solid symbols represent a descending sweep. The fluorescence drops to zero at a specific frequency, indicating the ejection of the ion from the trap. This point is marked by arrows for each isotope. The shaded areas highlight the operational frequency range for the removal of each respective isotope. Notably, after an ion's ejection, there is no recovery of fluorescence, signifying permanent removal from the trap, even as the modulation frequency is varied away from the resonant frequency.}
	\label{Fig3}
\end{figure}

To parametrically excite the trapped ion in the radial direction, we add a sinusoidal amplitude modulation on the rf voltage source. This renders the time-dependent rf voltage as $V_{\text{rf}}(t) = V_{0}[1+\epsilon_0\sin(\omega_m t)]\,\cos(\Omega t)$, where $\omega_m$ denotes the modulation frequency and $\epsilon_0$ is its fractional modulation amplitude.  It is crucial to note that when the modulation frequency is twice the secular frequency ($\omega_m = 2 \omega_r$), the parametric excitations experience resonant amplification. This leads to exponential growth in both the ion's motion and its energy.

In contrast, laser cooling damps the motion of trapped ion. This damping force can be expressed as $F = -\beta m v$, where $\beta = (4\hbar k^2/m) (\Delta/\Gamma) s/[1+s+4(\Delta/\Gamma)^2)]^2$. Here, $s$ is the saturation parameter and is proportional to the cooling laser's intensity. $\Delta$ refers to the cooling laser detuning and $\Gamma$ stands for the linewidth of the cooling transition. Incorporating laser cooling and parametric heating, the ion's equation of motion becomes
\begin{equation} 
\frac{d^2u}{dt^2}+\beta\frac{du}{dt}+ \frac{\Omega^2}{4}[a_u+2q'_u(t) \text{cos}(\Omega t)] u=0,
\label{EOM2}
\end{equation}
where $q'_u(t) = q_u[1+\epsilon_0\sin(\omega_m t)]$. If the heating overweighs the cooling, the ion's motional amplitude will increase until the ion is ejected from the trap. In contrast, if laser cooling dominates, its motional amplitude will diminish. To evaluate the competition, we numerically solve Eq.~(\ref{EOM2}) for various fractional amplitudes $\epsilon_0$ and cooling coefficients $\beta$. We obtain the assessment by comparing the initial and final motional amplitudes of a single ion after a brief propagation time $\Delta t =1~\text{ms}$. The results of this analysis are depicted in Fig. \ref{Fig2}.

The confining potential at the trap center can be approximated as a harmonic potential, known as the pseudopotential approximation. Using this approximation, the secular frequency can be expressed as $\omega_u' = (\Omega /2) \sqrt{a_u+(q_u'(t))^2/2}$, where $q_u'(t)$ is the time-dependent component created by the applied modulation. Consequently, the resulting confining potential in the radial direction can be derived as
\begin{equation}
U(r, t)=\frac{1}{2}m\omega_r'^2 r^2=\frac{1}{2}m\omega_r^2\left[ 1+ \eta(t)\right] r^2,
\label{mod_htrap}
\end{equation}
where $\omega_r = (\Omega /2) \sqrt{a_r+q_r^2/2}$ is the unperturbed secular frequency and $\eta(t)$ is a modulation term defined by $\eta(t) = \epsilon_0 q_r^2 \Omega^2 \sin(\omega_m t)/4\omega_r^2$.  The modulation in Eq.~(\ref{mod_htrap}) can be treated as a perturbation. Using the time-dependent perturbation method, one can compute the average transition rate between different oscillator states, and it becomes pronounced only when the modulation frequency is tuned near a parametric resonance, such as $\omega_m = 2\omega_r$. In this case, the oscillator's energy increases exponentially. The heating is characterized by the rate constant
\begin{equation}
\Gamma_{\text{heat}} = \frac{\pi}{2}\omega_r^2S(2\omega_r), 
\end{equation}
where $S(\omega)$ is the one-sided power spectrum of the modulation \cite{PhysRevA56R1095}. In the context of sinusoidal modulation, the one-sided power spectrum manifests nonzero values solely at the modulation frequency, specifically $S(\omega) = (\eta_0^2/2)\delta(2\omega_r-\omega)$ with $\eta_0$ representing the fractional modulation amplitude of the trap.

This parametric heating effect can be compared with laser cooling. The experimental cooling rate is $\Gamma_{\text{cool}} = 8590~\text{s}^{-1}$, and heating surpasses cooling when $S({2\omega_r})$ exceeds $2.95\times10^{-10}~\text{Hz}^{-1}$. A critical modulation amplitude of the trap can also be determined. Assuming the parametric resonance has a width of $\eta_0\omega_r$. An integration of the heating over this width, when compared with the integration from the cooling, leads to a critical modulation amplitude for the trap of $2.5\times10^{-3}$. Given that the trap's modulation is twice that of the RF voltage ($\eta_0 \approx 2\epsilon_0$), one finds the critical fractional modulation amplitude of the rf field $\epsilon_c = 1.25\times10^{-3}$.

\section*{Isotope-specific ion removal}\label{single_ions}

To investigate ion removal via parametric excitations, we start by preparing single ions of specific isotopes using the photoionization process. While preparing the ions of $^{40,42,44}\textrm{Ca}^{+}$ is straightforward, trapping $^{43}\textrm{Ca}^+$ ions poses a greater challenge due to their low abundance. When we observe the fluorescence from the targeted isotope, we terminate the ionization process and introduce a small amplitude modulation to the rf source that powers the ion trap. This modulation resonantly enhances parametric excitations when its frequency is approximately twice the secular frequency, leading to significant ion heating. We sweep the modulation frequency through the resonance in both ascending and descending directions to map out the boundaries of parametric resonances (see Fig.~\ref{Fig3}). To ensure our results are insensitive to variations in sweep rates, we adopt an extremely slow frequency sweep rate of approximately 1 kHz over 60 seconds, aimed at saturating the heating effect. Through this process, we identify distinct frequency windows suitable for ejecting specific isotopes of $\textrm{Ca}^{+}$.

According to parametric resonance theory, the full width of the resonance is determined by twice the product of the fractional amplitude and the oscillation frequency, leading to a value of 1.2 kHz under the experimental conditions described in Fig.~\ref{Fig3}. However, we have measured significantly larger widths, approximately 10 kHz, for all the isotopes. This notable deviation can come from the presence of technical factors that are not included in our simple model. These factors include fluctuations in the secular frequencies of the trap, which is proportional to the amplitude of the rf drive on the trap. For example, a 1\% deviation in our rf voltage $V_0$ leads to a 7.5 kHz deviation in the secular frequency, significantly larger than the predicted width. Furthermore, our data presented in Fig.~\ref{Fig3} is especially prone to this effect, since the measurement taken at each frequency usually lasts a few seconds. Despite observing larger widths, we can still identify the central frequency of each width as the resonant frequency for each isotope. These frequencies are in agreement with the theoretical estimates within an error of approximately 0.5\%.

\begin{figure}
	\centering
	\includegraphics[width=0.45\textwidth]{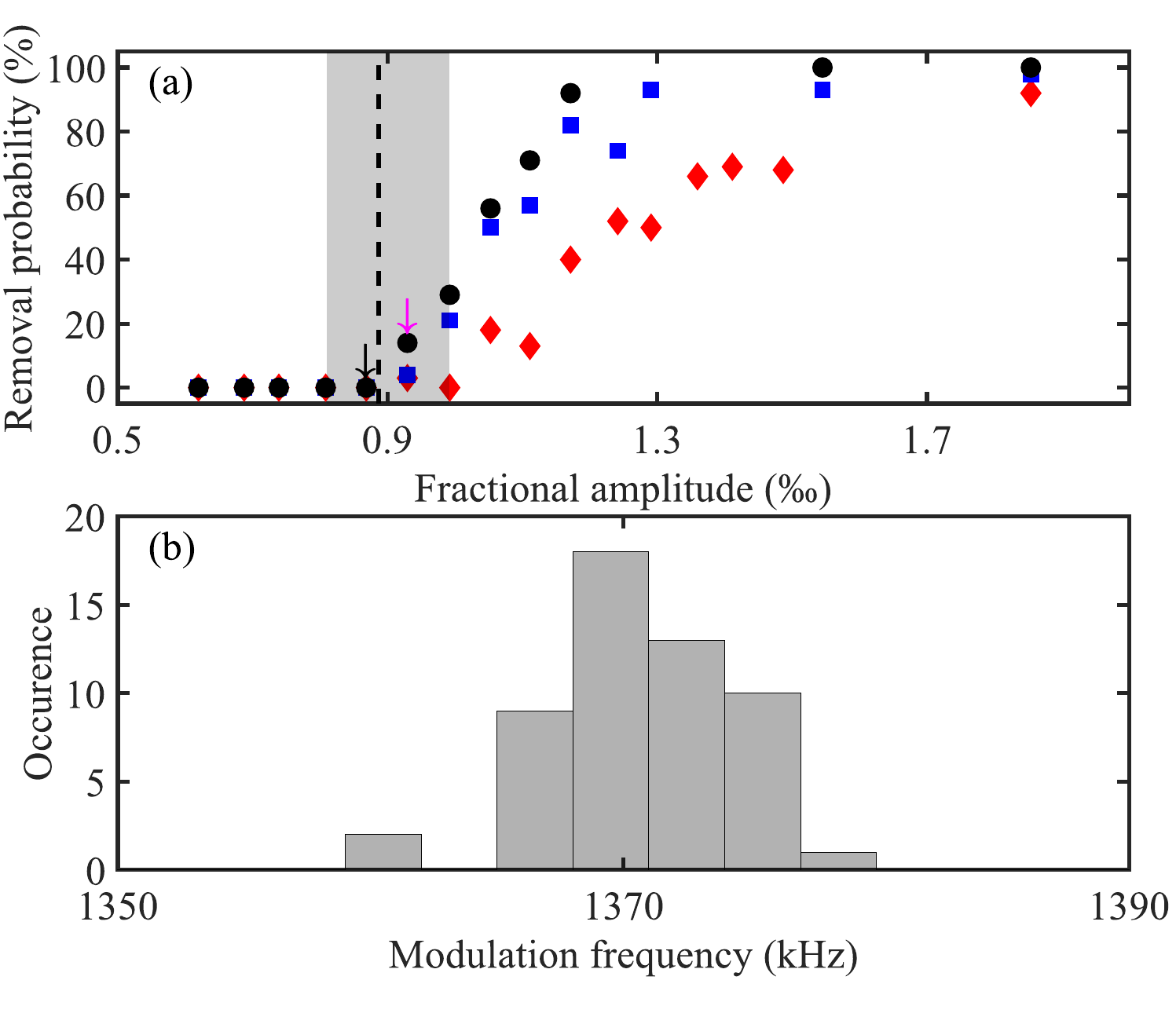}
	\caption{(a) Removal probability as a function of fractional amplitude. Each data point represents the removal outcome for 50 $^{40}\textrm{Ca}^+$ ions. Data were collected at three different frequency ramp speeds: 1 kHz/s (black circles), 2 kHz/s (blue squares), and 4 kHz/s (red diamonds). The dashed line indicates the fractional amplitude at which parametric heating equals laser cooling. This threshold was determined by solving Eq.~\ref{EOM2} using our experimental parameters, and the shaded region highlights its associated uncertainty. The two arrows indicate the data points used in Fig.~\ref{Fig2}. (b) Selective ion removal in a two-ion crystal of $^{43}$Ca$^+$ and $^{40}$Ca$^+$. The graph illustrates the frequency-dependent efficacy of $^{40}$Ca$^+$ removal. The most efficient removal frequency observed aligns with the results from single ion removal experiments.}
	\label{Fig4}
\end{figure}

To examine the interplay between parametric heating and laser cooling, we measure the ion removal probability's dependence on the fractional modulation amplitude $\epsilon_0$. For each fixed modulation amplitude, we sweep the modulation frequency across the parametric resonance and measure the resultant removal probability. This procedure is repeated for different modulation amplitudes.  The result is shown in Fig.~\ref{Fig4}(a). A sharp increase in removal probability was observed across three different ramp speeds when the fractional amplitude exceeded a certain threshold (0.886~$\text{\textperthousand}$). This critical point indicates where the effects of parametric heating surpass those of laser cooling.

Next, we analyze the ion removal process within a two-ion crystal. Our goal is to determine how the presence of additional ions in the trap affects the resonance frequency for the selective removal of a specific isotope. In these measurements, we laser-cool a $^{43}\textrm{Ca}^{+}$ ion, which in turn sympathetically cools a $^{40}\textrm{Ca}^{+}$ ion, forming a two-ion crystal. Once the ions are prepared, we activate the amplitude modulation and sweep its frequency across the 2$\omega_r$ resonance of $^{40}\textrm{Ca}^{+}$, a value previously determined and depicted in Fig.~\ref{Fig3}. Throughout the process, the fluorescence of the $^{43}\textrm{Ca}^{+}$ ion is monitored, using its position as an indicator for the successful ejection of $^{40}\textrm{Ca}^{+}$. We record the frequency at which the unwanted isotope is ejected. Fig.~\ref{Fig4}(b) illustrates the outcomes from 50 experimental repetitions. A comparison with Fig.~\ref{Fig3} reveals that the resonance frequency for $^{40}\textrm{Ca}^{+}$ removal matches those measured in the single-ion experiments. Importantly, the $^{43}\textrm{Ca}^{+}$ ions consistently remain trapped as the $^{40}\textrm{Ca}^{+}$ ions are removed. Afterwards, we extend these experiments to include various conditions, such as the presence of multiple $^{43}\textrm{Ca}^{+}$ and $^{40}\textrm{Ca}^{+}$ ions. In those measurements, we find that modulating the trap at the same resonance frequencies can still effectively remove $^{40}\textrm{Ca}^{+}$ without affecting the number of $^{43}\textrm{Ca}^{+}$.


The strong coupling between the two ions should, in principle, alter the resonance conditions. This effect, however, is mitigated when the ion crystal melts. In our experiments, we observed that even when the parametric drive is initially off-resonant for the two-ion system, with sufficient time, it can still effectively melt the two-ion crystal by utilizing the modulation frequency configured for single-ion resonance. This leads to the decoupling of the ions, at which point the parametric drive becomes increasingly efficient at heating and subsequently removing the unwanted isotope.

\section*{Preparation of $^{43}\textrm{Ca}^{+}$}\label{multiple_removal}
Building on our prior investigations of isotope purification, we have devised a scheme to prepare pure $^{43}\textrm{Ca}^{+}$ samples. Initially, we adjust the frequency of the 423-nm laser to resonate with the $4\text{S} \rightarrow 4\text{P}$ transition of $^{43}\textrm{Ca}$, aiming to maximize its ionization rate. Despite employing isotope-selective photoionization, the inadvertent production of unwanted isotopes remains inevitable due to overlapping absorption lines and significant disparity in abundance. 

We explore various procedures combining photoionization and resonant parametric excitations to prepare $^{43}\textrm{Ca}^{+}$. First, we tune the cooling lasers to cool $^{43}\textrm{Ca}^{+}$ ions from hot samples, and at the same time, we apply the resonant parametric excitations to remove the majority of the unwanted species, specifically $^{40}\textrm{Ca}^{+}$. This scheme is ineffective because the trapped $^{43}\textrm{Ca}^{+}$ ions can be continuously removed by the charge exchange collisions with neutral Ca atoms \cite{PhysRevA.69.042502}. In an effort to reduce these collisions, we opt for cooling $^{40}\textrm{Ca}^{+}$ instead of $^{43}\textrm{Ca}^{+}$; cooling the majority species yields a colder sample and hence suppresses the collision rate. 

\begin{figure}
	\centering
	\includegraphics[width=0.4\textwidth]{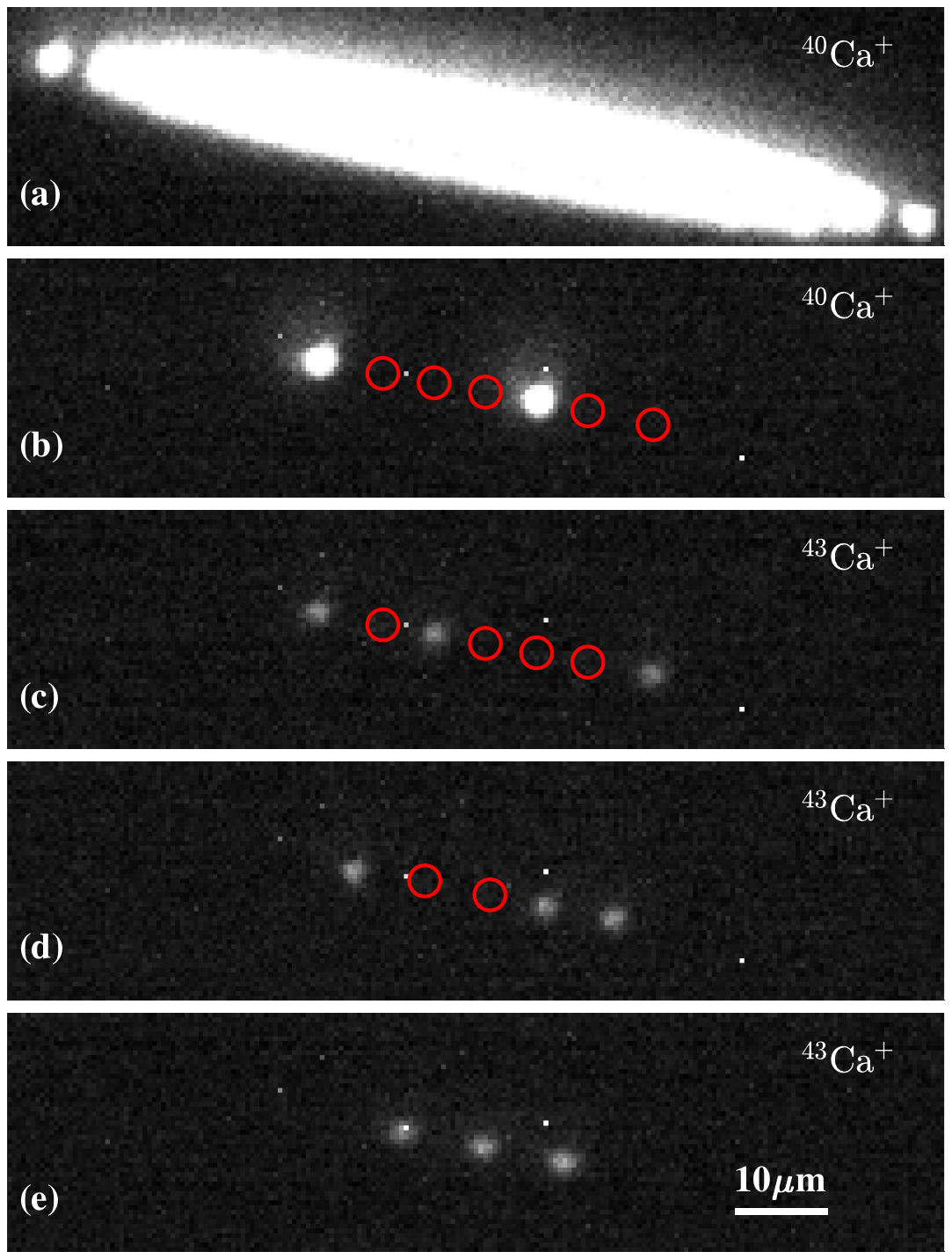}
	\caption{Fluorescence images illustrating the $^{43}\textrm{Ca}^{+}$ purification process within an isotope mixture: (a) Initial fluorescence reveals $^{40}\textrm{Ca}^{+}$ ions, the most abundant isotope in the mixture. (b) As purification commences, the majority of $^{40}\textrm{Ca}^{+}$ ions are removed. However, a few are left behind to assist in cooling other isotopes. Red circles mark the positions of dark ions. (c) Modifying the laser frequencies to specifically target $^{43}\textrm{Ca}^{+}$ results in three illuminated ions within the mixture. (d) Post the second removal stage, all $^{40}\textrm{Ca}^{+}$ ions are purged, bringing down the count of dark ions to two. (e) With the elimination of $^{44}\textrm{Ca}^{+}$ ions, the purification process centered on $^{43}\textrm{Ca}^{+}$ reaches completion.}
	\label{Fig5}
\end{figure}

In this revised scheme, we apply resonant parametric excitations post-loading, fine-tuning the modulation amplitude and duration to retain a few $^{40}\textrm{Ca}^{+}$ ions to cool the remaining dark ions in the trap. During the removal, ion crystals can be completely melted, significantly decreasing the fluorescence. Subsequently, the amplitude modulation is stopped, and we switch the lasers to cool $^{43}\textrm{Ca}^{+}$ ions. At this stage, we can determine the number of ions in the trap by analyzing the positions of the ions and the size of the crystal. After confirming the total number of ions, we reactivate the parametric excitations to remove all remaining dark ions. These unwanted ions typically consist of $^{40}\textrm{Ca}^{+}$, $^{42}\textrm{Ca}^{+}$, and $^{44}\textrm{Ca}^{+}$. We carefully sweep the amplitude modulation frequency across the parametric resonances specific to each unwanted isotope, ultimately isolating only $^{43}\textrm{Ca}^+$ ions in the trap. The whole procedure is depicted in Fig.~\ref{Fig5}.

\begin{figure}
	\centering
	\includegraphics[width=0.4\textwidth]{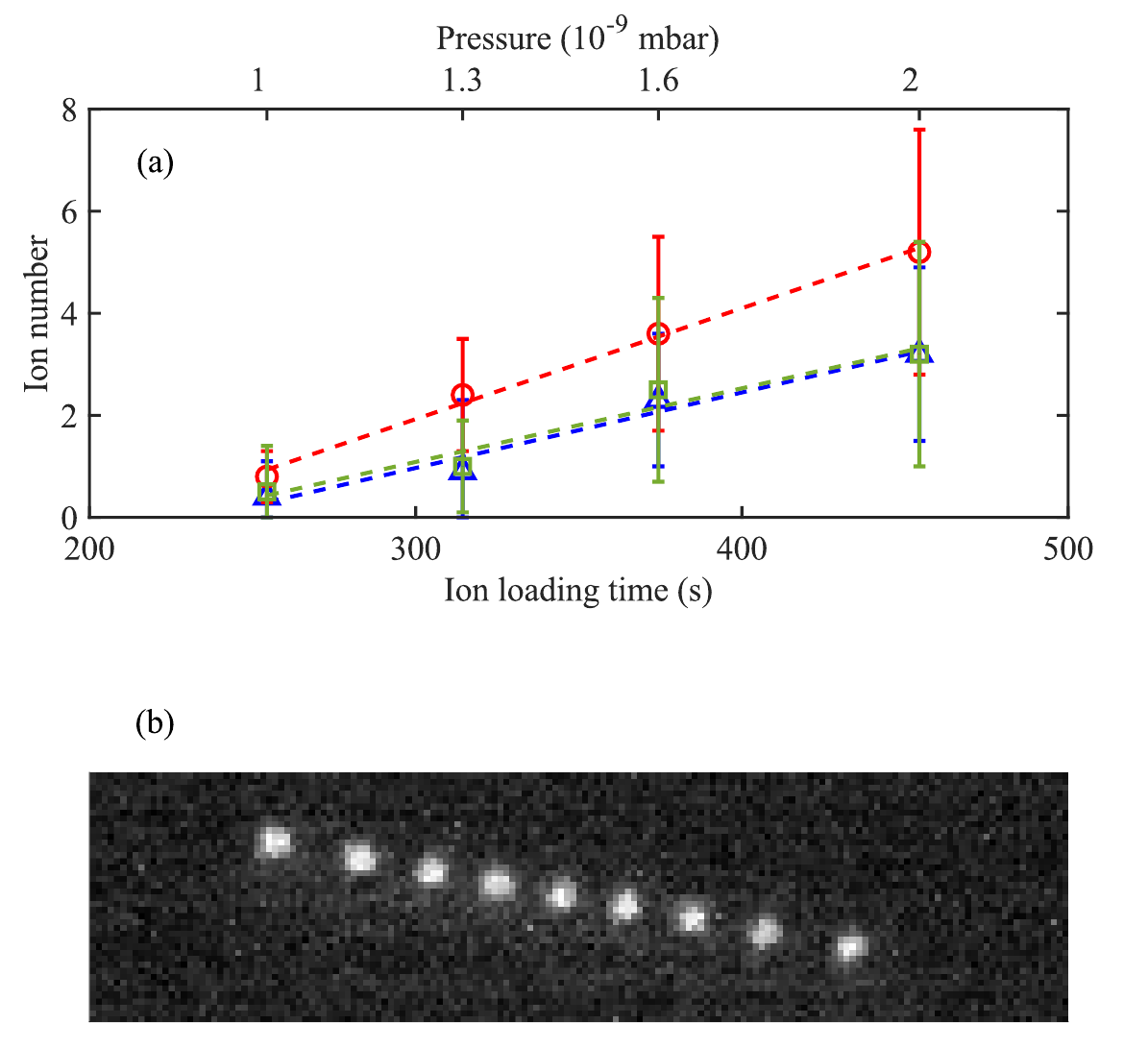}
	\caption{(a) Number of ions versus ionization time and background pressure for $^{43}\textrm{Ca}^{+}$ (circle), $^{42}\textrm{Ca}^{+}$ (triangle), and $^{44}\textrm{Ca}^{+}$ (square). The respective ratios of $^{43}\textrm{Ca}^{+}$ to $^{44}\textrm{Ca}^{+}$ and $^{43}\textrm{Ca}^{+}$ to $^{42}\textrm{Ca}^{+}$ are 1.51 and 1.47. The dashed lines are linear fits to the data. (b) Fluorescence image displaying a pure sample of $^{43}\textrm{Ca}^+$ ions. The image distinctly presents nine individual $^{43}\textrm{Ca}^+$ ions, highlighting the efficiency of the purification process within the ion trap.}
	\label{Fig6}
\end{figure}


In addition, we control the duration of ionization to regulate the production of $^{43}\textrm{Ca}^+$ ions. In our experiment, it is challenging to stabilize the atomic flux from the effusive oven, and it is equally difficult to assess the oven temperature variations. Despite these challenges, we have made efforts to keep all experimental parameters constant, varying only the time for ionization. This approach ensures that the flux passing through the ionizing regime increases over time, thereby increasing the number of $^{43}\textrm{Ca}^+$ ions generated. Fig.~\ref{Fig6}(a) depicts the ion counts for various isotopes, observed after removing unwanted isotopes, as a function of the ionization time. Although this method introduces considerable fluctuations in the $^{43}\textrm{Ca}^+$ ion count, it suggests that generating a substantial or target number of $^{43}\textrm{Ca}^+$ ions with adequate loading time is feasible, as demonstrated in Fig.~\ref{Fig6}(b).

\section*{Conclusion}\label{conclusion}
In summary, we have successfully demonstrated the preparation of $^{43}\textrm{Ca}^{+}$ ions in an ion trap using photoionization and resonant parametric excitations, showing the method's ability to purify a rare isotope. Notably, the purification technique provides the necessary mass resolution for selectively removing any specific calcium isotopes. This capability allows us to extract undesired isotopes individually without affecting the target calcium ions. Combining with a miniaturized ion trap, we aim to explore new possibilities for quantum sensing. We also investigate the interplay between parametric heating and laser cooling, finding a good agreement between our theoretical calculations and experimental results. The models enable us to predict the modulation amplitude required for isotope purification.  





 








\section*{Acknowledgements}

We thank Ite A. Yu and Y.-W. Liu, and L.-B. Wang for their valuable discussions and early contributions to the project. This work was supported by the Ministry of Education of Taiwan and the National Science and Technology Council of Taiwan. S.T. acknowledges additional support from the Multidisciplinary and Competitive Programs for Higher Education Sprout Project.

\section*{Author contributions statement}


C.K., Y.H., C.J., Y.C. and S.T. conducted the experiments and analyzed the results.  C.K. and S.T. wrote the manuscript. S.T. supervised the research.

\section*{Data availability statement}

The datasets used and analyzed during the current study are available from the corresponding author on reasonable request.







\end{document}